\documentclass[fleqn,twoside]{article}
\usepackage{espcrc2}
\usepackage{amsmath,amssymb,latexsym}
\usepackage{graphicx}

\newcommand{\be}{\begin{equation}}
\newcommand{\ee}{\end{equation}}

\newcommand{\bea}{\begin{eqnarray}}
\newcommand{\eea}{\end{eqnarray}}
\renewcommand{\deg}{^{\circ}}

\title{Lattice approach to high--energy hadron--hadron scattering}

\author{M. Giordano\address[a]{Dipartimento di Fisica, Universit\`a di Pisa,
    and INFN, Sezione di Pisa\\  
    Largo Pontecorvo 3, I--56127 Pisa, Italy.}\thanks{Speaker at the conference.}
  and E. Meggiolaro\addressmark[a]}

\begin{document}
\thispagestyle{empty}

\begin{abstract}
We discuss the non perturbative approach to the problem of high--energy
hadron--hadron (dipole--dipole) scattering at low momentum transfer
by means of numerical simulations in Lattice Gauge Theory.
\end{abstract}

\maketitle

\section{Introduction}

The prediction from f\mbox{}irst principles of total cross sections at high energy
is one of the oldest open problems of hadronic physics.
Present--day experimental data are well described by 
a universal {\it pomeron}--like power--law behaviour (see, for
example, Ref.~\cite{pomeron-book} and references therein),
$\displaystyle \sigma_{\rm tot}^{(hh)} (s) \mathop{\sim}_{s \to \infty}
\left( {s/s_0} \right)^{\epsilon_P}$ ,
where the so--called {\it soft pomeron} intercept is $\epsilon_P \simeq 0.08$,
but this is forbidden as a true asymptotic behaviour by the
well--known Froissart--Lukaszuk--Martin theorem~\cite{FLM}.
As we believe QCD to be the fundamental theory of strong
interactions, it should predict the correct asymptotic behaviour;
nevertheless, a satisfactory explanation is still lacking.

The problem of total cross sections is part of the more
general problem of high--energy %{\it elastic} 
scattering at low
transferred momentum, the so--called {\it soft high--energy
  scattering}. As soft high--energy processes possess two dif\mbox{}ferent
energy scales, the total center--of--mass energy squared 
$s$ and the transferred momentum squared $t$,
smaller than the typical energy scale of strong
interactions ($|t| \lesssim 1~ {\rm GeV}^2 \ll s$), we cannot fully rely on
perturbation theory. A genuine non perturbative
approach in the framework of
QCD has been proposed by Nachtmann  in~\cite{Nachtmann91}, and further
developed in~\cite{DFK,Nachtmann97,BN,Dosch,LLCM1}: using a functional
integral approach, high--energy hadron--hadron elastic scattering
amplitudes are shown to be governed by the correlation function of
certain Wilson loops def\mbox{}ined in Minkowski space. Moreover, as it has
been shown in~\cite{Meggiolaro02,Meggiolaro05}, such a correlation
function can be reconstructed by analytic continuation from its
Euclidean counterpart, i.e., the correlation function of two Euclidean
Wilson loops, that can be calculated using the non perturbative
methods of Euclidean Field Theory.

In~\cite{lat} we have investigated this problem 
by means of numerical simulations in Lattice Gauge Theory (LGT). Although 
we cannot obtain an analytic expression in this way, nevertheless
this is a f\mbox{}irst--principle approach that provides (inside the errors)
the true QCD expectation for the relevant correlation function. %% ,
In this contribution,
after a quick survey of the non perturbative approach to soft
high--energy scattering in the case of meson--meson {\it elastic} scattering,
we will present our numerical approach based on LGT, and we will show how
the numerical results can be compared to the existing analytic models.

\section{Meson--meson elastic scattering amplitudes and Wilson--loop correlators}

We sketch here the non perturbative approach to soft high--energy scattering;
see~\cite{lat} for a more detailed presentation. As it has been shown
in~\cite{DFK,Nachtmann97,BN}, the elastic 
meson--meson scattering amplitude can be reconstructed from the
scattering amplitude of two $q\bar{q}$ colour dipoles, after
averaging over the transverse sizes and the longitudinal momentum
fractions of the dipoles. The central quantity in this approach is 
a certain (properly normalised) correlation function (in the sense of the QCD
functional integral) of two Wilson loops in the fundamental
representation, def\mbox{}ined in Minkowski space--time, running along the paths
made up of the quark and antiquark straight--line classical
trajectories and closed at proper times $\pm T$ by straight--line
paths in the transverse plane. 

In~\cite{Meggiolaro02,Meggiolaro05}
(see also~\cite{crossing,Meggiolaro07}) it has been shown that, under
certain analyticity hypotheses, this correlation function can be
reconstructed from the Euclidean correlation function of two Euclidean
Wilson loops, $\widetilde{\cal W}_1$ and $\widetilde{\cal W}_2$,  
\be
{\cal G}_E(\theta;T;\vec{z}_\perp;1,2) \equiv 
{ \langle \widetilde{\cal W}^{(T)}_1 \widetilde{\cal W}^{(T)}_2 \rangle \over
\langle \widetilde{\cal W}^{(T)}_1 \rangle
\langle \widetilde{\cal W}^{(T)}_2 \rangle } - 1
\label{GE}
\ee
(here ``1'' and ``2'' stand respectively for $\vec{R}_{1\perp},f_1$ and
$\vec{R}_{2\perp},f_2$); the two loops run along two rectangular paths
$\widetilde{\cal C}_1$ and $\widetilde{\cal C}_2$, made up of the ``Euclidean
trajectories'' of the partons,
\bea
\lefteqn{\widetilde{\cal C}_1 :
X^{1q[\bar{q}]}_{E}(\tau)
 = z + {p_{1E} \over m} \tau + 
f^{q[\bar{q}]}_1 R_{1E} , }\nonumber\\
\lefteqn{\widetilde{\cal C}_2 :
X^{2q[\bar{q}]}_{E}(\tau)
 = {p_{2E} \over m} \tau + 
f^{q[\bar{q}]}_2 R_{2E}}
\label{traj}
\eea
where $p_{1[2]E} =
m \left( [-]\sin{\theta \over 2}, \vec{0}_\perp, \cos{\theta \over 2} \right)$,
$R_{iE} = (0,\vec{R}_{i\perp},0)$, $f^{q}_i = 1-f_i$, $f^{\bar{q}}_i = -f_i$
($i=1,2$), $z_E = (0,\vec{z}_\perp,0)$, with
$\vec{R}_{i\perp}$ and $f_i$ the transverse sizes and longitudinal
momentum fractions of the two dipoles, and $\vec{z}_\perp$ the
impact--parameter distance between the two loops in the transverse plane. The
paths are closed at proper times $\pm T$ by straight--line paths in the
transverse plane; here $T$ acts 
as an IR cutof\mbox{}f which has to be removed in the end. As the elastic
scattering amplitude of two meson states is expected to be an
IR--f\mbox{}inite physical quantity~\cite{BL}, we expect the limit
$T\to\infty$ to be f\mbox{}inite, so that we can def\mbox{}ine $\displaystyle {\cal
  C}_E(\theta;\vec{z}_\perp;1,2)\equiv {\cal
G}_E(\theta;T\to\infty;\vec{z}_\perp;1,2)$. 
 Note that ${\cal G}_E$ is a real function, as can
be shown making use of the charge--conjugation invariance of the
functional integral. 

Finally, the {\it meson--meson} scattering amplitudes can be written
as 
\bea
\lefteqn{{\cal M}_{(hh)} (s,t;1,2) = -i~2s~\displaystyle\int d^2 \vec{z}_\perp e^{i \vec{q}_\perp \cdot
   \vec{z}_\perp}} \nonumber \\
& &  \times \displaystyle\int d1~|\psi_1(1)|^2  \displaystyle\int
 d2~|\psi_2(2)|^2 \nonumber \\
& & \times~~{\cal C}_M (%\chi;
\theta \to -i\log \left( {s/m^2} \right);
\vec{z}_\perp;1,2);
\eea
here $\vec{q}_{\perp}$ is the (transverse) transferred momentum
($t=-|\vec{q}_{\perp}|^2$), and $\psi_1$ and $\psi_2$ are the wave
functions which describe the two interacting mesons. Total cross sections are
then recovered via the optical theorem.

In the following we will set for simplicity $f_1=f_2=1/2$, which is known to
be a good approximation for hadron--hadron interactions~\cite{pomeron-book,Dosch}.

\section{Wilson--loop correlators on the lattice}

The gauge--invariant Wilson--loop correlation function
${\cal G}_E$ is a natural candidate for a lattice computation, but
some care has to be taken due to the explicit breaking of $O(4)$
invariance on a lattice. As straight lines on a lattice can be either parallel
or orthogonal, we are forced to use {\it of\mbox{}f--axis} Wilson loops to
cover a signif\mbox{}icantly large set of angles. To stay as close as
possible to the continuum case, the loop sides  are evaluated on the
lattice paths that minimise the distance from the true, 
continuum paths: this can be easily accomplished making use of the well--known {\it Bresenham
algorithm}~\cite{Bres} to f\mbox{}ind the required  ``{\it minimal distance
paths}'' corresponding to the sides of the loops.
The relevant Wilson loops  $\widetilde{\cal
  W}_L(\vec{l}_{\parallel};\vec{r}_{\perp};n)$ are then characterised
by the position $n$ of their center and by two two--dimensional vectors
$\vec{l}_{\parallel}$ and $\vec{r}_{\perp}$, corresponding
respectively to the longitudinal and transverse sides of the loop.

On the lattice we then def\mbox{}ine the correlator 
\bea
\label{eq:corr_lat}
\lefteqn{{\cal
  G}_L(\vec{l}_{1\parallel},\vec{l}_{2\parallel};\vec{d}_{\perp};\vec{r}_{1\perp},\vec{r}_{2\perp})
 } \nonumber \\
 & & \equiv \frac{\langle\widetilde{\cal 
    W}_L(\vec{l}_{1\parallel};\vec{r}_{1\perp};d)\widetilde{\cal
    W}_L(\vec{l}_{2\parallel};\vec{r}_{2\perp};0)\rangle}{\langle\widetilde{\cal 
  W}_L(\vec{l}_{1\parallel};\vec{r}_{1\perp};d)\rangle\langle\widetilde{\cal
  W}_L(\vec{l}_{2\parallel};\vec{r}_{2\perp};0)\rangle} - 1,
\eea
where $d=(0,\vec{d}_{\perp},0)$, $\vec{d}_{\perp} = (d_2,d_3)$, and
moreover
\bea
\lefteqn{{\cal
  C}_L(\hat{l}_{1\parallel},\hat{l}_{2\parallel};\vec{d}_{\perp};\vec{r}_{1\perp},\vec{r}_{2\perp})}
  \nonumber \\
& & \equiv \lim_{L_1,L_2\to\infty} {\cal
  G}_L(\vec{l}_{1\parallel},\vec{l}_{2\parallel};\vec{d}_{\perp};\vec{r}_{1\perp},\vec{r}_{2\perp}),
\eea
where $L_i \equiv |\vec{l}_{i\parallel}|$ are def\mbox{}ined to be the lengths of the
longitudinal sides of the loops in lattice units,  and
$\vec{l}_{i\parallel}\equiv L_i\hat{l}_{i\parallel}$. 
In the continuum limit, where $O(4)$ invariance is restored, we expect  
\bea
\label{eq:contlim}
\lefteqn{{\cal
  C}_L(\hat{l}_{1\parallel},\hat{l}_{2\parallel};\vec{d}_{\perp};\vec{r}_{1\perp},\vec{r}_{2\perp})}
  \nonumber\\
& &  \mathop\simeq_{a\to 0} {\cal
  C}_E(\theta;a\vec{d}_{\perp};a\vec{r}_{1\perp},1/2,a\vec{r}_{2\perp},1/2),
\eea
where $\hat{l}_{1\parallel}\cdot\hat{l}_{2\parallel} \equiv
\cos\theta$ def\mbox{}ines the relative angle $\theta$, and $a$ is the
lattice spacing. 

\section{Numerical results}
\begin{figure}[t]
\resizebox{75.1mm}{!}{\includegraphics{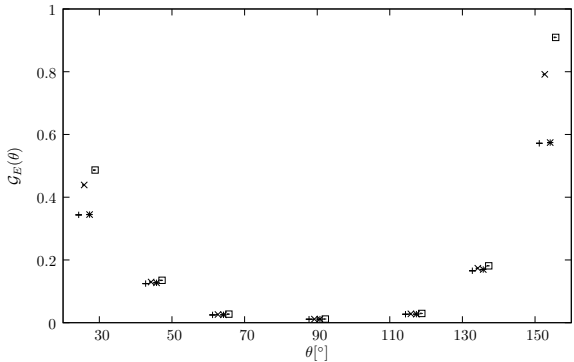}}
\vspace{-35pt}
\caption{Lattice data plotted against $\theta$ for various lengths of
  the loops.}
\label{fig:1}
\vspace{-14pt}
\end{figure}
As already pointed out in the Introduction,
numerical simulations cannot provide the analytic expression for the
relevant correlation function, but nevertheless, as these simulations are
f\mbox{}irst--principles calculations, 
they provide the ``correct'' (inside the errors) prediction of QCD.
Approximate analytic calculations have then to be compared with the
lattice data, in order to test the goodness of the approximations
involved. In particular, we are interested in the dependence on the
relative angle $\theta$, as it encodes the energy dependence of the
scattering amplitudes, which is recovered after the proper analytic
continuation. 

In Fig.~\ref{fig:1} we show, as an example, the lattice data for
${\cal G}_L$ in the case of
parallel transverse sides with
$|\vec{r}_{1\perp}|=|\vec{r}_{2\perp}|=1$ at $d=0$, plotted against the angle 
$\theta$ for various lengths of the loops.
These data are obtained using Wilson action for $SU(3)$ pure--gauge ({\it
  quenched}\/) theory, on a  $16^4$ lattice at $\beta=6.0$. The data are quite
stable against variations of the lengths, so that we can take the 
largest--length data as a reasonable approximation of ${\cal C}_L$.

\begin{figure}[t]
\resizebox{75.1mm}{!}{\includegraphics{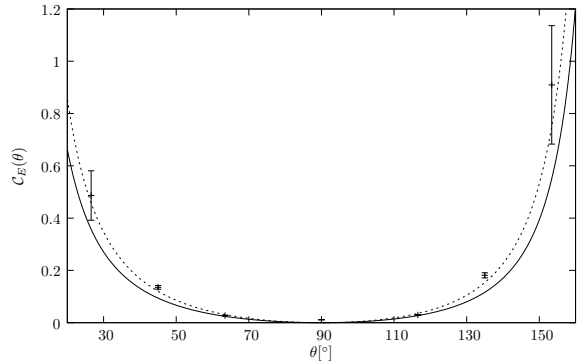}}
\vspace{-35pt}
\caption{Comparison of lattice data with the SVM prediction (solid) and
  with a best--f\mbox{}it with the SVM functional form (dotted).}
\label{fig:2}
\vspace{-13pt}
\end{figure}
\begin{figure}[thb]
\resizebox{75.1mm}{!}{\includegraphics{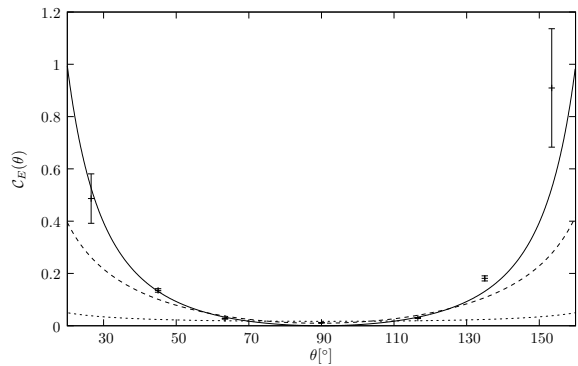}}
\vspace{-35pt}
\caption{Comparison of lattice data with best--f\mbox{}its with the
  lowest--order perturbative (solid), the ILM (dotted) and the AdS/CFT
  (dashed) expressions.} 
\label{fig:3}
\vspace{-14pt}
\end{figure}
In Figs.~\ref{fig:2} and~\ref{fig:3} we compare ${\cal C}_L$ with
the prediction of various models (the loop conf\mbox{}iguration is the same
as in Fig.~\ref{fig:1}). While the {\it Stochastic Vacuum Model}
(SVM)~\cite{LLCM2} provides a fully quantitative prediction, that can be 
directly compared with the data, the {\it Instanton Liquid Model}
(ILM)~\cite{instanton1} and the AdS/CFT correspondence~\cite{JP1} give  
only the qualitative dependence on the angle $\theta$, so that a
comparison can be made by trying to f\mbox{}it the data with the given
functional form. In Fig.~\ref{fig:2} we show the SVM prediction, together
with a best--f\mbox{}it with the SVM functional form; in Fig.~\ref{fig:3} we show
the best--f\mbox{}its with the expressions obtained in perturbation 
theory to lowest order~\cite{BB,LLCM2,Meggiolaro05}, in the ILM and
using the AdS/CFT correspondence. 
A detailed discussion of the results is given in~\cite{lat}; here we
simply note that the agreement of the numerical data with the various
models is not fully satisfactory, and further investigations have to
be made, both on the numerical and on the analytical side. 
\begin{figure}[t]
\resizebox{75.1mm}{!}{\includegraphics{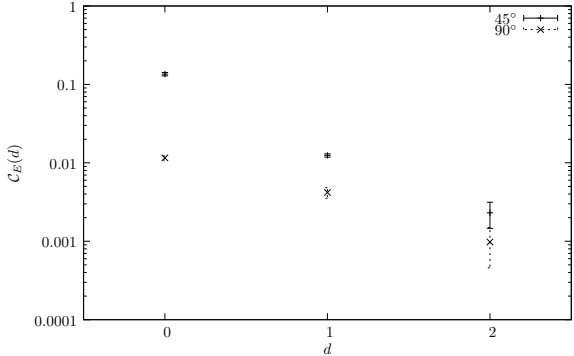}}
\vspace{-35pt}
\caption{Dependence of ${\cal C}_E$ on the distance
  for $\theta=45\deg$ and $\theta=90\deg$ in the case
  $\vec{r}_{1\perp}=\vec{r}_{2\perp}$, $|\vec{r}_{1\perp}|=1$,
  $\vec{r}_{1\perp}\parallel \vec{d}_{\perp}$ (logarithmic scale).} 
\label{fig:4}
\vspace{-13pt}
\end{figure}
We want also to remark that while perturbative ef\mbox{}fects seem to be dominant at
short distances between the loops, non perturbative ef\mbox{}fects are already relevant at
distances of about $0.2$ fm; however, as the correlation function is
rapidly decreasing with the distance $d=|\vec{d}_{\perp}|$ between the centers of the loops
(see Fig.~\ref{fig:4}), a detailed study at large distances is dif\mbox{}f\mbox{}icult,
and requires the use of noise reduction techniques.

\begin{figure}[t]
\resizebox{75.1mm}{!}{\includegraphics{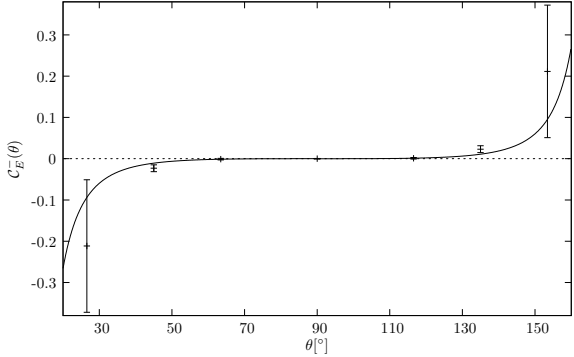}}
\vspace{-35pt}
\caption{``Antisymmetric'' part of lattice data, and corresponding SVM
  prediction.}
\label{fig:5}
\vspace{-14pt}
\end{figure}
Lattice data show also the presence of {\it odderon} contributions to
dipole--dipole scattering. Indeed, as explained
in~\cite{Meggiolaro07,lat}, making use of the {\it crossing--symmetry}
relations for loops~\cite{crossing} one can show that the {\it
  crossing--odd} component of the dipole--dipole scattering amplitudes
is related via the usual analytic continuation to the
antisymmetric (with respect to $\pi/2$) component ${\cal C}_E^-$ of 
${\cal C}_E$: this quantity is shown in Fig.~\ref{fig:5}, together with the
coresponding SVM prediction (the loop conf\mbox{}iguration is the same as in
Fig.~\ref{fig:1}).

\end{document}